\documentstyle[floats,epsf,eqsecnum,prl,aps]{revtex}
\begin{document}
\draft
\lefthyphenmin=2
\righthyphenmin=3

\title{\rightline{\rm{FERMILAB-Pub-98/237-E}} 
\vskip 0.3 cm
Small Angle $J/\psi $ Production in $p\overline{p}$ Collisions at 
$\sqrt{s}$= 1.8~TeV}
% LIST_OF_AUTHORS.TEX                 7/10/98            
%
\author{                                                                      
%% names begin here                                                           
B.~Abbott,$^{40}$                                                             
M.~Abolins,$^{37}$                                                            
V.~Abramov,$^{15}$                                                            
B.S.~Acharya,$^{8}$                                                           
I.~Adam,$^{39}$                                                               
D.L.~Adams,$^{48}$                                                            
M.~Adams,$^{24}$                                                              
S.~Ahn,$^{23}$                                                                
H.~Aihara,$^{17}$                                                             
G.A.~Alves,$^{2}$
% ...
\\                                                                            
\vskip 0.70cm                                                                 
\centerline{(D\O\ Collaboration)}                                             
\vskip 0.70cm                                                                 
}                                                                             
\address{                                                                     
\centerline{$^{1}$Universidad de Buenos Aires, Buenos Aires, Argentina}       
\centerline{$^{2}$LAFEX, Centro Brasileiro de Pesquisas F{\'\i}sicas,         
                  Rio de Janeiro, Brazil}                                     
\centerline{$^{3}$Universidade do Estado do Rio de Janeiro,                   
                  Rio de Janeiro, Brazil}                                     
\centerline{$^{4}$Universidad de los Andes, Bogot\'{a}, Colombia}             
\centerline{$^{5}$DAPNIA/Service de Physique des Particules, CEA, Saclay,     
                  France}                                                     
\centerline{$^{6}$Panjab University, Chandigarh, India}                       
\centerline{$^{7}$Delhi University, Delhi, India}                             
\centerline{$^{8}$Tata Institute of Fundamental Research, Mumbai, India}      
\centerline{$^{9}$Kyungsung University, Pusan, Korea}                         
\centerline{$^{10}$Seoul National University, Seoul, Korea}                   
\centerline{$^{11}$CINVESTAV, Mexico City, Mexico}                            
\centerline{$^{12}$Institute of Nuclear Physics, Krak\'ow, Poland}            
\centerline{$^{13}$Institute for Theoretical and Experimental Physics,        
                   Moscow, Russia}                                            
\centerline{$^{14}$Moscow State University, Moscow, Russia}                   
\centerline{$^{15}$Institute for High Energy Physics, Protvino, Russia}       
\centerline{$^{16}$University of Arizona, Tucson, Arizona 85721}              
\centerline{$^{17}$Lawrence Berkeley National Laboratory and University of    
                   California, Berkeley, California 94720}                    
\centerline{$^{18}$University of California, Davis, California 95616}         
\centerline{$^{19}$University of California, Irvine, California 92697}        
\centerline{$^{20}$University of California, Riverside, California 92521}     
\centerline{$^{21}$Florida State University, Tallahassee, Florida 32306}      
\centerline{$^{22}$University of Hawaii, Honolulu, Hawaii 96822}              
\centerline{$^{23}$Fermi National Accelerator Laboratory, Batavia,            
                   Illinois 60510}                                            
\centerline{$^{24}$University of Illinois at Chicago, Chicago,                
                   Illinois 60607}                                            
\centerline{$^{25}$Northern Illinois University, DeKalb, Illinois 60115}      
\centerline{$^{26}$Northwestern University, Evanston, Illinois 60208}         
\centerline{$^{27}$Indiana University, Bloomington, Indiana 47405}            
\centerline{$^{28}$University of Notre Dame, Notre Dame, Indiana 46556}       
\centerline{$^{29}$Purdue University, West Lafayette, Indiana 47907}          
\centerline{$^{30}$Iowa State University, Ames, Iowa 50011}                   
\centerline{$^{31}$University of Kansas, Lawrence, Kansas 66045}              
\centerline{$^{32}$Louisiana Tech University, Ruston, Louisiana 71272}        
\centerline{$^{33}$University of Maryland, College Park, Maryland 20742}      
\centerline{$^{34}$Boston University, Boston, Massachusetts 02215}            
\centerline{$^{35}$Northeastern University, Boston, Massachusetts 02115}      
\centerline{$^{36}$University of Michigan, Ann Arbor, Michigan 48109}         
\centerline{$^{37}$Michigan State University, East Lansing, Michigan 48824}   
\centerline{$^{38}$University of Nebraska, Lincoln, Nebraska 68588}           
\centerline{$^{39}$Columbia University, New York, New York 10027}             
\centerline{$^{40}$New York University, New York, New York 10003}             
\centerline{$^{41}$University of Rochester, Rochester, New York 14627}        
\centerline{$^{42}$State University of New York, Stony Brook,                 
                   New York 11794}                                            
\centerline{$^{43}$Brookhaven National Laboratory, Upton, New York 11973}     
\centerline{$^{44}$University of Oklahoma, Norman, Oklahoma 73019}            
\centerline{$^{45}$Brown University, Providence, Rhode Island 02912}          
\centerline{$^{46}$University of Texas, Arlington, Texas 76019}               
\centerline{$^{47}$Texas A\&M University, College Station, Texas 77843}       
\centerline{$^{48}$Rice University, Houston, Texas 77005}                     
}                                                                             
%end                                                                          
\date{\today}
\maketitle

\begin{abstract}
{
This paper presents the first  measurement  of the  inclusive $J/\psi $ production 
cross section in the forward pseudorapidity region $2.5\leq | \eta^{J/\psi} | \leq 3.7$ 
in $p\overline{p}$ collisions at $\sqrt{s} = 1.8~$TeV. The results are
based on 9.8 pb$^{-1}$ of data collected using the D\O\ detector at the Fermilab Tevatron 
Collider. The inclusive
$J/\psi $ cross section for transverse momenta between 1 and 16 GeV/$c$ is
compared with  theoretical models of charmonium production.
}
\end{abstract}


%\narrowtext
\twocolumn
In high energy $p\overline{p}$ collisions $J/\psi$'s are produced 
directly, from decays of higher mass charmonium states 
[$\chi$ and $\psi(2S)$], and from $b$~quark decays.
Existing experimental results in the central rapidity
region from UA1\cite{1} at $\sqrt{s} = 0.63~$TeV, and from CDF\cite{2} and D\O\cite{3} 
at $\sqrt{s} = 1.8~$TeV demonstrate that then  
measured inclusive $J/\psi $ transverse momentum distribution 
cannot be described solely by contributions from $b~$quark decays and 
prompt production predicted by the color
singlet model\cite{4}. 
In the color singlet model the charmonium meson retains the quantum numbers of the 
produced $c\overline{c}$ pair and thus each $J/\psi$ state can only be directly produced via 
the corresponding hard scattering color singlet subprocess. The model predicts   
direct $J/\psi$ and $\psi(2S)$ production rates fifty times
smaller than those observed by CDF\cite{2}. To explain 
this discrepancy, a color octet model was introduced\cite{5,5a,cho}. The color octet 
mechanism extends the color singlet approach by taking into account the production of 
$c\overline{c}$  pairs in a color octet configuration accompanied by a gluon. The color octet 
state evolves into a color singlet state via emission of a soft gluon. The parameters of the 
model  were derived from a fit to CDF data for    direct $J/\psi$ and 
$\psi(2S)$ production at  central rapidity.
In this article we utilize the large rapidity
coverage of the D\O\ muon system to study the process 
$p\overline{p}\rightarrow J/\psi \ +X\rightarrow \mu ^{+}\mu ^{-}+X$ 
in  previously unexplored kinematical regions  of small $J/\psi$  transverse momenta 
and large rapidities. We compare our results with theoretical predictions 
extended into this  kinematic  domain.

The D\O\ detector\cite{6} consists of three main systems:
central and forward drift chambers, used to identify charged tracks for  pseudorapidity 
$|\eta|\le 3.2$;
the uranium-liquid argon calorimeter with nearly hermetic coverage for $|\eta| \le 4$; 
and the muon system. The detector component most relevant to this analysis is the
Small Angle MUon Spectrometer (SAMUS)\cite{7,8} consisting of 
magnetized iron toroids and drift tube stations on each side of the interaction region with 
pseudorapidity coverage of  2.2 $< |\eta^{\mu}| < $3.3 for a single muon.

The SAMUS stations, three in each arm, consist of three planes of 29 mm diameter 
drift tubes: vertical, horizontal, and inclined at $45^{\circ}$.   The list  of  
tubes containing hits is sent to the trigger system   and drift times are used
for  offline track reconstruction. Muon track reconstruction is based on a  Kalman fit\cite{kal}  
to the three-dimensional coordinates of muons passing through the SAMUS stations,
one before the toroidal magnet and two after, and the coordinates of the event vertex. 
The  muon 
momentum resolution $\sigma_p/p$    is about  20\%,   limited by  SAMUS coordinate
resolution  and by Coulomb scattering in the calorimeter and muon toroid.

The data were collected using   the multilevel trigger system.  
The Level 0 trigger\cite{11}  is used to select  single interaction
events with hits in scintillator hodoscopes situated on both sides of the interaction region. 
At  the Level 1\cite{12},   signals from 
individual SAMUS tubes are OR'ed to provide 12 cm wide hodoscopic elements. The 
trigger requires a pattern of hits in the trigger elements
consistent with at least one muon with transverse momentum 
$p_T^{\mu} > 3$~GeV/$c$ coming 
from the interaction region.
Due to high tube occupancy $ (\simeq 4\%)$  by soft electrons and 
positrons, we implement a ``multiplicity cut'' at the Level 1 trigger. This cut rejects an
event if the number of hit trigger elements in a vertically oriented tube plane exceeds a 
fixed threshold.

The 
logic of the Level 1.5 trigger is similar to that of the Level 1 trigger,
but is based on better spatial segmentation (1.5 cm vs. 12 cm).
Events which pass the Level 1.5 trigger  are  digitized and
sent to the Level 2 software trigger implemented on a farm of  VAX
stations, where reconstruction of muon tracks without using drift times is performed.   
The calorimeter information is used in the Level 2 trigger to 
confirm the muon through its energy deposition.

Even with the multiplicity cut, the counting rates of the Level 1 and Level
1.5  triggers are high in comparison with the allocated trigger
bandwidth. To further reduce counting rates, we use prescales up to 10
for the dimuon trigger.

In the offline analysis, we select events with one interaction vertex, a single muon or 
dimuon trigger, and at least two reconstructed muon tracks.  
Each muon candidate is required to have at least 15 hits on a track out of an average of 18. 
The energy deposition in the cells of the hadronic 
calorimeter along the muon track is required to exceed 1.5~GeV, and to be spread 
contiguously among all 
five calorimeter layers.  To ensure a good momentum measurement, we require 
$p^{\mu} \leq  150$~GeV/$c$ and a minimum traverse magnetic  field integral of 
1.2~T$\cdot$m.
In total, 1779  events  with  opposite sign muon pairs and 281 events with 
same sign muon pairs are selected from the data sample with 
integrated luminosity of $9.8\pm0.5~$pb$^{-1}$\cite{17}.  The estimated fraction of 
background tracks from accidental hit combinations in the final data sample is below 1\%.

\begin{figure}[t]
\epsfxsize=3.4in\epsfbox{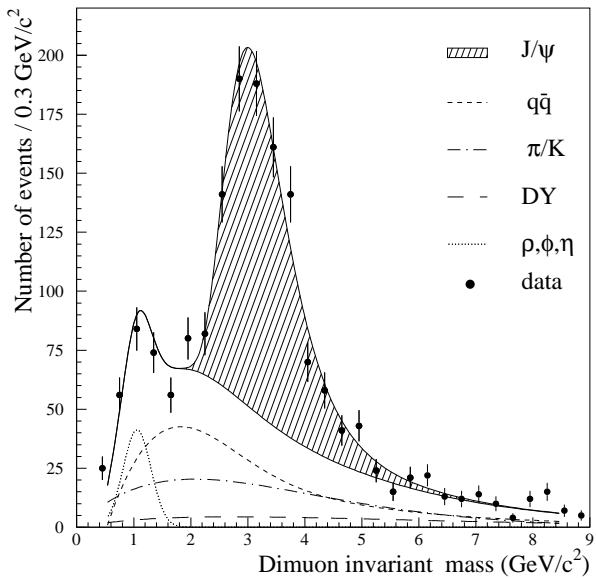}
\caption{ The invariant mass spectrum of opposite sign dimuons with 1.0 
$\leq p_{T}^{\mu \mu }\leq 16$~GeV/$c$ and 2.5 $\leq |\eta ^{\mu\mu }| \leq $
3.7. The hatched area indicates the $J/\psi $ signal above the sum of the
backgrounds.}
\label{fig1}
\end{figure}

The opposite sign dimuon invariant mass distribution $M_{\mu\mu}$ for events with transverse
momentum in the range  $1.0\leq p_{T}^{\mu \mu }\leq 16$~GeV/$c$  
and  pseudorapidity 2.5 $\leq |\eta ^{\mu\mu }| \leq 3.7$
is shown in Fig.~\ref{fig1}. In addition to the $J/\psi$ signal, 
other contributions to the dimuon  spectrum with $M_{\mu\mu}< 9$~GeV/$c^2$ are 
expected to  come from
$b\overline{b}$ and $c\overline{c}$ production (jointly denoted as $q\overline{q}$) with  the 
heavy quarks decaying 
semileptonically or via sequential semileptonic decays, Drell-Yan production (DY), 
decay of light mesons (e.g. $\rho$, $\phi$, $\eta$), and $\pi$ or  $K$ decays.

To estimate the background and simulate the $J/\psi$ detection efficiency, we use a 
sample of Monte Carlo (MC) events from the {\footnotesize  PYTHIA 5.7}\cite{pyt} and 
 {\footnotesize JETSET}\cite{jet} MC generators for each of the dimuon processes mentioned 
above,  except $\pi$ and  $K$ decays.   The $J/\psi$ events are generated 
(assuming no $J/\psi$ polarization) with $p_T^{J/\psi}$  
from 1~GeV/$c$ to 20~GeV/$c$ and $|\eta ^{J/\psi }|$  between 2.0 and 4.0. Generated
dimuon events are simulated using  {\footnotesize D\O\ GEANT}\cite{gea} and mixed with 
minimum bias events  from the data to simulate the combinatoric background. These
simulated dimuon events are then subjected to a full trigger simulation and processed  
with the standard D\O\ reconstruction program. 

Based on MC studies we approximate the $J/\psi$ signal by a Gaussian 
function of $1/M_{\mu\mu}$
to account for limited muon momentum resolution.
The mass spectrum in Fig.~\ref{fig1}  is fit by the sum of  the $J/\psi$  
signal (with the width and mean value as free parameters) and MC mass distributions 
for background processes (with free normalization). The number of events due to 
$\pi$ and $K$ decays is    estimated from the data using like-sign dimuon  events.
The fit yields $691\pm41$ $J/\psi$ events with mean mass 
$\langle M_{\mu\mu} \rangle = 3.03\pm 0.03$~GeV/$c^2,$ and 
standard deviation $ \sigma_M=0.56 \pm 0.03$~GeV/$c^2$.

The  dimuon  mass resolution does not allow
a clear separation of the  $J/\psi$  and $\psi (2S)$  states.
The fit of the invariant mass distribution yields a 90\% C.L. upper limit of the $\psi(2S)$ 
fraction
in the signal associated with the $J/\psi $   of 15\%.
A direct measurement of inclusive $\psi (2S)$ production for $|\eta ^{\psi }|<0.6$ 
performed by CDF\cite{2}
shows that the $J/\psi $ differential cross section is approximately 13 times
larger than that of the $\psi (2S)$. 

The inclusive differential cross section of $J/\psi $ production is
calculated from:
\[
\frac{d^{2}\sigma \left( \left\langle p_{T}^{i}\right\rangle ,\left\langle
|\eta ^{j}|\right\rangle \right) }{dp_{T}^{i}d\eta ^{j}}=\frac{1}{%
L\varepsilon _{ij}}\frac{ N_{ij}}{\Delta p_{T}^{i}\Delta \eta ^{j}},
\]
where $L$ is the total integrated luminosity, $\varepsilon _{ij}$ is the  $J/\psi$  detection
efficiency,  and $N_{ij}$ is the number of $J/\psi $ events in the 
$\Delta p_{T}^{i}$, $\Delta \eta ^{j}$ interval.

To  calculate  the number of $J/\psi$ events, the fit to the mass spectrum
is performed in five  $\eta^{\mu\mu} $ and nine $p_T^{\mu\mu}$
intervals. To reduce  the  errors of
the fit in the high $p_T^{\mu\mu}$ bins, the $p_T^{\mu\mu}$
dependence of the fraction of events attributed to  $J/\psi $ 
is fit to a linear function and the
results of this fit are used to obtain the number of $J/\psi$ events.
\begin{figure}[t]
\epsfxsize=3.4in\epsfysize=4.in\epsfbox{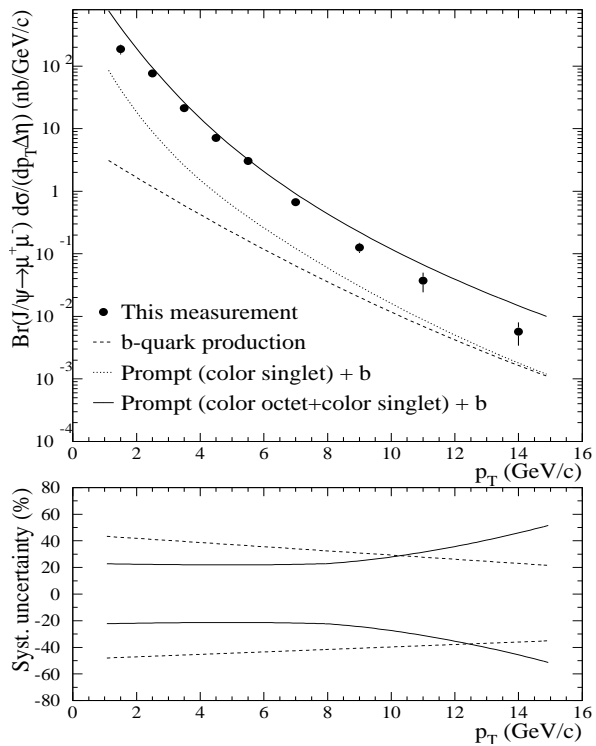}
\caption{The $p_{T}$ dependence of the  $J/\psi$ differential cross 
section  and its  
theoretical predictions (upper figure). Only the statistical errors 
are shown. The lower figure presents systematic uncertainties; the solid curves 
are the sum of all systematic errors, the dashed curves represent the uncertainty band  due 
to $J/\psi$ polarization. 
The upper (lower) dashed curve corresponds to 100$\%$ transverse (longitudinal)
polarization.}
\label{fig2}
\end{figure}

\mediumtext

\begin{table*}
\caption{$J/\psi $ inclusive differential cross sections 
$Br(J/\psi\rightarrow \mu ^{+}\mu ^{-})d^{2}\sigma /dp_{T}d\eta$  (nb/GeV/$c$).}
\begin{tabular}{cp{0.1in}ccccc}
& \multicolumn{1}{r}{$\eta^{J/\psi}$}
& 2.5 - 3.7 
& 2.65 
& 2.95
& 3.25
& 3.55 \\
%\hline
$p_{T}^{J/\psi}$ (GeV/$c$) &&&&&& \\ \hline
1.5 && 187 $\pm$ 34~ & ~--- & 137 $\pm$ 55~ & 183 $\pm$ 57~ & 130 $\pm$ 45~ \\ 
2.5 && 77 $\pm$ 8~ & ~--- & 69 $\pm$ 18 & 69.9 $\pm$ 8.1~ & 45.7 $\pm$ 5.3~ \\ 
3.5 && 21.3 $\pm$ 2.1~ & 21.4 $\pm$ 6.4~ & 23.8 $\pm$ 2.8~ & 17.6 $\pm$ 2.0~ & 
15.6 $\pm$ 1.7~ \\ 
4.5 && 7.14 $\pm$ 0.77 & 7.9 $\pm$ 1.7 & 8.06 $\pm$ 0.83 & 5.59 $\pm$ 0.86 & 
5.40 $\pm$ 0.83 \\ 
5.5 && 3.03 $\pm$ 0.36 & 4.20 $\pm$ 0.65 & 3.22 $\pm$ 0.36 & 2.44 $\pm$ 0.46
& --- \\ 
7.0 && 0.667 $\pm$ 0.075 & 1.07 $\pm$ 0.13 & 0.758 $\pm$ 0.076 & 0.77 $\pm$ %
0.16 & --- \\ 
9.0 && 0.126 $\pm$ 0.023 & 0.212 $\pm$ 0.033 & 0.132 $\pm$ 0.021 & 0.131 $\pm$ %
0.039 & ---\\ 
11.0 && 0.037 $\pm$ 0.013 & 0.086 $\pm$ 0.020 & 0.050 $\pm$ 0.014 & ---
& --- \\ 
14.0 && 0.0057 $\pm$ 0.0023 & 0.0206 $\pm$ 0.0039 & 0.0064 $\pm$ 0.0020 & ---
& ---
\end{tabular}
\label{tab2}
\end{table*}
\narrowtext

The efficiency of  $	J/\psi $ detection includes acceptance,
trigger efficiency, reconstruction efficiency, and offline cuts and is given by 
\[
\varepsilon _{ij}=\frac{N\left( p_{T}^{i},\eta ^{j}\right) \cdot \varepsilon
_{\tt cor}}{N_{\tt tot}\left( p_{T}^{i},\eta ^{j}\right) },
\]
where $N\left( p_{T}^{i},\eta ^{j}\right)$ is the number of events in a given $p_T^i$, $\eta^j$  bin
which passed all selection criteria, $N_{\tt tot}\left( p_{T}^{i},\eta ^{j}\right) $ is the total
number of generated events in a bin, and $\varepsilon_{\tt cor}$ 
is the correction factor for effects not simulated in MC. 
The $\varepsilon _{\tt cor}$ includes efficiencies for the
Level 2 trigger calorimeter confirmation of (91 $\pm$ 2)\% for the dimuon trigger 
and (95 $\pm$ 1)\% for the single muon trigger, and for offline cuts  not simulated by the MC of
(79 $\pm$ 4)\%. 
Efficiencies for those cuts are obtained from the data and include
$(88 \pm 1)\%$ for the single vertex cut,  $(94 \pm 3)\%$  for the energy deposition
cut, and  $(96 \pm 2)\%$ for the cut on the number of hits on a track.

The measured $J/\psi$  spectrum is unfolded  
to correct for the  momentum and pseudorapidity smearing  using 
the technique of  Ref.~\cite{20}.
The  correction factors vary from
1.7 at low $p_T^{J/\psi}$ to  0.4
for  $p_T^{J/\psi}>8$~GeV/$c$.

The calculated differential cross section is
fit to an exponential function. The results  are used for
interpolation of the cross sections from average values of  $\left\langle
p_{T}^{i}\right\rangle $ and $\left\langle |\eta^{j}|\right\rangle $ to the
centers of the selected intervals.
The inclusive differential $J/\psi$ cross section  averaged over a
rapidity range of $2.5 \leq |\eta ^{J/\psi }| \leq 3.7$ is shown in Fig.~\ref{fig2}.  
Results  for finer rapidity bins are collected in Table~\ref{tab2}. 
The uncertainties quoted there are statistical only.
\begin{table}[h]
\caption{Systematic errors of the $J/\psi $ cross sections.}
\label{tab3}
\begin{tabular}{cc}
Source & Systematic Error \\ 
\hline
Unfolding procedure & 15\% \\ 
$J/\psi $ background determination &   7.2\% -- 30\% \\ 
$J/\psi$ detection efficiency   & 7\% \\
Level 1 multiplicity cut  & 6\% \\
Total integrated luminosity & 5.4\% \\ 
$\psi (2S)$ contamination & $ _{-5\%}^{+0\%}$ \\ 
$d^{2}\sigma /dp_{T}d\eta$  total &  20\%  -- 36\%\\
\hline
Averaging over $2.5<|\eta^{J/\psi}|<3.7$  &  5\% -- 30\% \\ 
$d\sigma /dp_{T}\Delta\eta$  total  &  21\% -- 47\%
\end{tabular}
\end{table}

The largest $(>2\%)$   systematic uncertainties are summarized in
Table~\ref{tab3}.  
The  contribution   from  the unfolding  is derived  from comparison with
the  bin-by-bin unfolding technique\cite{zhi}. 
The uncertainties  in the determination of the background  and averaging cross 
section over the SAMUS pseudorapidity acceptance  vary for different $p_T^{J/\psi}$  
and are caused by uncertainties in the  parameterization of the data.    
The difference in the parameters of measured and generated $J/\psi$ mass distributions 
as well as the accuracy of the spectrometer description in  the detector simulation  
are used to estimate the $J/\psi$ detection efficiency uncertainty. The uncertainty due to the  
Level 1 
multiplicity cut was determined by varying the threshold of this cut by one trigger element.   
The  results in  Table \ref{tab2} are obtained for the case  of zero $J/\psi$ polarization. 
The additional uncertainty up to $^{+40\%}_{-45\%}$ due to possible $J/\psi$
polarization is shown in Fig.~\ref{fig2}  along with the $p_T^{J/\psi}$ dependence of the total 
systematic error.

\begin{figure}[t]
\epsfxsize=3.4in\epsfbox{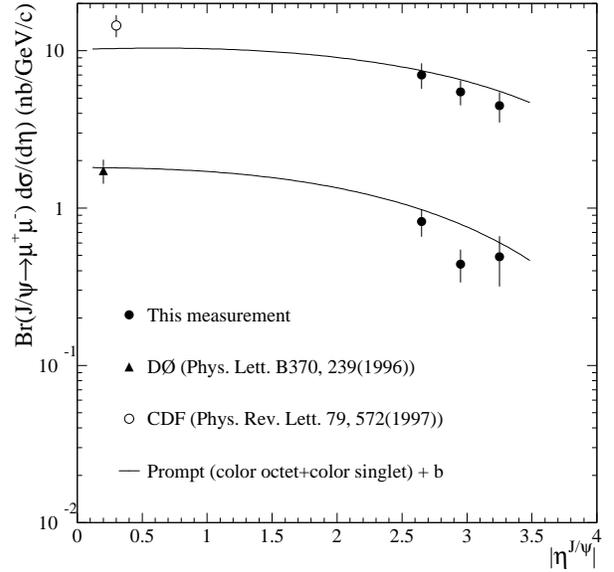}
\caption{The pseudorapidity dependence of the  $J/\psi $ production
cross section with $p_T> 5$~GeV/$c$ (upper points and curve) and $p_T>
8$~GeV/$c$ (lower points and curve). The error bars are statistical and systematic errors
(polarization uncertainties not included) summed in quadrature.}
\label{fig3}
\end{figure}

In Fig.~\ref{fig2} we compare the $J/\psi$ cross section with  current models of 
charmonium production.  For $J/\psi$ from  $b$~quarks 
we use the NLO QCD predictions\cite{26} with the renormalization/factorization
scale $\mu= \frac{1}{3} \sqrt{m_b^2+{p_T^b}^2}$, where $m_b$ and $p_T^b$ are the parent
 $b$~quark 
mass and transverse momentum, respectively.  The scale is chosen to match theory 
predictions to the 
published D\O\  $b$~quark cross sections in the central rapidity region\cite{3}. 
We use {\footnotesize ISAJET}\cite{isa} to fragment $b$~quarks into $J/\psi$.
The color octet and color singlet contributions  to the direct $J/\psi$ production
and  radiative $\chi$ decays are 
taken from Ref.~\cite{cho}. The term representing  the direct $J/\psi$ production
is increased by 12\% to account for the contribution from $\psi(2S)$ decays\cite{2}.

Fig.~\ref{fig3} shows 
the pseudorapidity dependence of the measured $J/\psi$  cross section 
for $p_T^{J/\psi}>5$ and 8 GeV/$c$ along with the corresponding central rapidity 
measurements of  D\O\ \cite{3} and CDF\cite{2}.
Within uncertainties, the  color octet model plus $b$~quark decays
describe the $\eta$ dependence of the inclusive $J/\psi$ production in the  full rapidity region.

In conclusion, we have made the first measurement of  inclusive $J/\psi$ production 
in the forward rapidity region $2.5<|\eta^{J/\psi}|<3.7$ in $p\overline{p}$ collisions  at 
$\sqrt{s} = 1.8~$TeV. The data show   good agreement with the  theoretical  
predictions based on $b$~quark decays and the color octet model of 
direct charmonium production.

% Acknowledgement_paragraph.tex
%
We thank the staffs at Fermilab and collaborating institutions for their
contributions to this work, and acknowledge support from the 
Department of Energy and National Science Foundation (U.S.A.),  
Commissariat  \` a L'Energie Atomique (France), 
Ministry for Science and Technology and Ministry for Atomic 
   Energy (Russia),
CAPES and CNPq (Brazil),
Departments of Atomic Energy and Science and Education (India),
Colciencias (Colombia),
CONACyT (Mexico),
Ministry of Education and KOSEF (Korea),
and CONICET and UBACyT (Argentina).


\begin{references}

% LIST_OF_VISITOR_ADDRESSES.TEX                            06/18/97
%
\bibitem[*]{ecuador}
Visitor from Universidad San Francisco de Quito, Quito, Ecuador.

\bibitem[\dag]{beijing}
Visitor from IHEP, Beijing, China.
\vskip 0.25cm
\bibitem{1}    C. Albajar {\it et al.}  (UA1 Collaboration), Phys. Lett. B {\bf  256}, 112 (1991).
\bibitem{2}    F. Abe {\it et al.} (CDF Collaboration), Phys. Rev. Lett. {\bf  69}, 3704 (1992);
{\sl ibid} {\bf 79}, 572 (1997); 
{\sl ibid} {\bf 79}, 578 (1997). 
\bibitem{3} S. Abachi {\it et al.} (D\O\ Collaboration), Phys. Lett. B {\bf  370}, 239 (1996).
\bibitem{4} R. Baier and R. Ruckl, Z. Phys. C {\bf 19}, 251 (1983).
\bibitem{5} E. Braaten and S. Fleming, Phys. Rev. Lett. {\bf  74}, 3327 (1995);
 P. Cho and M. Wise, Phys. Lett. B {\bf 346}, 129 (1995).
\bibitem{5a} M. Cacciari {\it et al.}, Phys. Lett. B {\bf 356}, 553 (1995).
\bibitem{cho} P. Cho and A.K. Leibovich, Phys. Rev. D {\bf 53}, 6203 (1996).
The expected cross section for direct $J/\psi$  production and the contribution from 
$\chi$ decays   were calculated using  code provided by P. Cho.
\bibitem{6} S. Abachi {\it et al.}  (D\O\ Collaboration), 
Nucl. Instrum. Methods Phys. Res., Sect. A {\bf 338}, 185 (1994).
\bibitem{7}  C. Brown {\it et al.},  Nucl. Instrum. Methods Phys. Res., Sect.  
A {\bf 279}, 331 (1989).
\bibitem{8}  Yu. Antipov {\it et al.}, Nucl. Instrum. Methods Phys. Res., Sect. 
A {\bf 297}, 121 (1990).
\bibitem{kal} P. Billoir, Nucl. Instrum. Methods Phys. Res., Sect.  A {\bf 225}, 352 (1984).
\bibitem{11}  J. Bantly {\it et al.}, IEEE Trans on Nucl. Sci. {\bf  41}, 1274, (1994).
\bibitem{12} M. Abolins {\it et al.}, Nucl. Instrum. Methods Phys. Res., Sect. 
A {\bf  289}, 543 (1990);
  M. Fortner {\it et al.}, IEEE Transactions on Nuclear Science {\bf  38},
480 (1991).
\bibitem{17} J. Bantly {\it et al.}, Fermilab Technical Memo FNAL-TM-1995, 1997 
(unpublished).
\bibitem{pyt} T. Sj\"{o}strand, Computer Physics Commun. {\bf 39}, 347 (1986);
 T. Sj\"{o}strand and M. Bengtsson, Computer Physics Commun. {\bf 43}, 367 (1987).
\bibitem{jet} M. Bengtsson and T. Sj\"{o}strand,  Computer Physics Commun. {\bf 46}, 43 
(1987).
\bibitem{gea} R. Brun and F. Carminati, CERN Program Library Long Writeup W5013, 1993 
(unpublished); we use {\footnotesize GEANT} v3.15.
\bibitem{20} G. D'Agostini,   Nucl. Instrum. Methods Phys. Res., Sect.  A {\bf  362}, 487 (1995).
\bibitem{zhi} V.B. Anikeev, A.A. Spiridonov, and V.P. Zhigunov, 
 Nucl. Instrum. Methods Phys. Res., Sect. 
A {\bf 322}, 280 (1992).
\bibitem{26} The expected inclusive $b$ quark cross section was calculated using the MNR
code, provided by M. Mangano.
\bibitem{isa} F. Paige and S. Protopopescu, BNL Report No. 38304, 1986 (unpublished).


\end{references}
\end{document}